\documentclass[prl,twocolumn,showpacs]{revtex4}
\usepackage{graphicx,amsfonts,amssymb,amsmath}

\begin{document}

\title{Diffusion-Induced Oscillations of Extended Defects}

\date{\today}

\author{Alexander L. Korzhenevskii}
\affiliation{Institute for Problems of Mechanical Engineering,
RAS, Bol'shoi prosp. V. O., 61, St Petersburg, 199178, Russia}
\author{Richard Bausch}
\affiliation{Institut f{\"u}r Theoretische Physik IV,
 Heinrich-Heine-Universit{\"a}t D{\"u}sseldorf, Universit{\"a}tsstrasse
 1, D-40225 D{\"u}sseldorf, Germany}
\author{Rudi Schmitz}
\affiliation{Institut f{\"u}r Theoretische Physik A, RWTH Aachen University,
Templergraben 55, D-52056 Aachen, Germany}

\date{\today}

\begin{abstract}
From a simple model for the driven motion of a planar interface under the influence of a diffusion field we derive a damped nonlinear oscillator equation for the interface position. Inside an unstable regime, where the damping term is negative, we find limit-cycle solutions, describing an oscillatory propagation of the interface. In case of a growing solidification front this offers a transparent scenario for the formation of solute bands in binary alloys, and, taking into account the Mullins-Sekerka instability, of banded structures.
\end{abstract}


\maketitle

The interaction of propagating extended defects with a diffusion field frequently leads to oscillations or jerky motions of the defects. A prime example of such an effect is the oscillation of a solidification front, induced by the diffusion of the solute component in a dilute binary alloy, which is growing in the setup of  directional solidification. In a large number of metallic materials this leads to the formation of banded structures \cite{CGZK}, reflecting a periodic array of layers with high and low solute concentrations where the former ones show a dendritic microstructure. The appearance of similar banding effects has recently been discussed \cite{EP} in rapid solidification of colloids.

Layered structures are also generated by the oscillatory nucleation of a solid phase under the action of a diffusion field \cite{KCGOSI}. A related phenomenon is the oscillatory zoning, observed in solid solutions \cite{PDP} and in natural minerals \cite{SF}. Another notable scenario is that of diffusion-controlled jerky motions of a driven grain boundary \cite{Cahn}. A similar behavior of dislocations in metallic alloys leads to the Portevin-Le Chatelier effect \cite{LBEK}, denoting the appearance of jerky plastic deformations. We, finally, mention the oscillatory motion of a crack tip, which is coated by the nucleus of a new phase \cite{AA}, replacing the attached cloud of a diffusion field.

Theoretical discussions of such effects either are of a phenomenological type, like those in Ref. \cite{LBEK}, and partly in Refs. \cite{CGZK} and \cite{EP}, or they rely on a Fokker-Planck \cite{Cahn}, or a diffusion equation with non-equilibrium boundary conditions \cite{KS}. In all approaches the source of oscillatory defect motions is identified as an unstable regime where a reduction of the driving force leads to an increase of the defect velocity. Additional information is provided by simulations, based on phase-field models for directional solidification \cite{Conti} and for nucleation \cite{KCGOSI} processes.

In the present Letter we introduce an extremely simple but powerful model for the diffusion-induced          oscillatory motion of a planar interface, using the language adapted to the directional solidification of a dilute binary alloy. A major advantage of our approach is that it allows a transparent and, to a large extend, analytical evaluation. This includes a readjustment of the stability analysis by Merchant and Davis \cite{MD} who discovered an oscillatory instability, similar to that, discussed earlier by Coriell and Sekerka \cite{CS}. Also included is a clarifying analysis of the so far barely understood low-velocity sections of the cyclic trajectories, identified by Carrard et al. \cite{CGZK}, and by Karma and Sarkissian \cite{KS}. The limit-cycle behavior, describing the oscillations of the interface deep inside the unstable regime, is, finally, in remarkable agreement with the simulation results by Conti \cite{Conti}. 

Due to the restriction to a planar interface, our model is a one-dimensional version of the capillary-wave model, derived in Ref. \cite{PRE} from a phase-field model. It is given in dimensionless form by the set of equations

\begin{eqnarray}\label{model}
&&H=\frac{\gamma}{2}\int_{-\infty}^{+\infty}dz\,[C(z,t)-U(z-Z(t))]^2\,\,,\\
&&\partial_t Z=p\,\left(F-\frac{\delta H}{\delta Z}\right)\,\,\,,\,\,\,
\partial_t C=\partial_z^2\,\frac{1}{\gamma}\,\frac{\delta H}{\delta C}\,\,,\nonumber\\
&&F=F_P-m^2[Z(t)-v_P\,t]\nonumber
\end{eqnarray}
for the interface position $Z(t)$, and for the excess solute concentration $C(z,t)$ relative to its value $C_S\equiv 1$ in the solid phase. The parameter $\gamma$ measures the miscibility gap $\Delta C=C_L-C_S$ where $C_L$ is the solute concentration in the liquid phase, and $p$ measures the mobility of the interface. 
From the equilibrium condition $\delta H/\delta C=0$ it follows that $U(z-Z)$ is the equilibrium-concentration profile at some fixed temperature $T_S$. This profile reveals the smooth solid-liquid transition region of the original phase-field model, and is regarded as an input quantity of the model (\ref{model}). It effectively comprises non-equilibrium effects of sharp-interface descriptions, which are crucial for the behavior in the rapid-growth regime, including the solute-trapping effect \cite{AB}.

The driving force $F$ includes two quantities, appearing in the simplest scenario of directional solidification. One of them is a constant temperature gradient $S$, entering the parameter $m^2\equiv A\,\xi S/T_M$ where, in physical units, $\xi$ measures the width of the interface, visible in the profile $U(\zeta)$, $T_M$ is the melting temperature of the pure solvent, and $A$ is a numerical pre-factor of order one. Adopting from Ref. \cite{Conti} typical values for $S,T_M,\xi$, the parameter $m^2$ is of order $10^{-5}$. An independent second quantity is the velocity $v_P$, applied in pulling the growing crystal in opposite direction to the temperature gradient. The local temperature $T_P$ at the steady-state position $Z(t)=v_P\,t$, finally, determines the fragment $F_P\equiv A(T_S-T_P)/T_M$ of the driving force $F$.

The resulting equations for $Z(t)$ and $C(z,t)$ read

\begin{eqnarray}\label{integro}
&&\frac{1}{p}\,\dot Z(t)=F_P-m^2[Z(t)-v_P\,t]\\ &&-\,\gamma\int_{-\infty}^{+\infty}dz\,U'(z-Z(t))[C(z,t)-U(z-Z(t))]\,\,,\nonumber\\\,\nonumber\\
&&(\partial_t-\partial_z^2)C(z,t)=-\,U''(z-Z(t))\,\,.\nonumber
\end{eqnarray}
For steady-state boundary conditions $C(z=\pm\infty)=0$, they have the stationary solutions $Z(t)=v_P\,t$, and $C(z,t)=C_P(z-v_P\,t;v_P)$, resulting in the relations 

\begin{eqnarray}\label{balance}
&&\frac{1}{p}\,v_P=F_P+G_P(v_P)-G_P(0)\,\,,\\&&G_P(v_P)\equiv-\,\gamma\int_{-\infty}^{+\infty}d\zeta\,
U'(\zeta)\,C_P(\zeta;v_P)\,\,,\nonumber\\&&C_P(\zeta;v_P)=\int_{-\infty}^{\zeta}d\zeta'\,U'(\zeta')\,
\exp{[v_P(\zeta'-\zeta)]}\,\,.\nonumber
\end{eqnarray}

We are primarily interested in the late-stage behavior of non-stationary solutions $Z(t)$, and, therefore, look for a solution $C(z,t)$ of the last equation in Eqs. (\ref{integro}), obeying the boundary condition $C(z,-\infty)=0$. This leads to the expression

\begin{eqnarray}\label{concentration}
C(z,t)=\int_{-\infty}^{t}dt'\int_{-\infty}^{+\infty}dz'\,\partial_{z'}\mathcal{G}(z-z',t-t')&&\nonumber\\
\cdot\,U'(z'-Z(t'))\,\,,&&
\end{eqnarray}
involving the Green function 

\begin{equation}\label{Green}
\mathcal{G}(z,t)=\int_{-\infty}^{+\infty}\frac{dk}{2\pi}\,\exp{(-k^2t+ik\,z)}\,\,.
\end{equation}
After the substitutions $\zeta\equiv z-Z(t),\zeta'\equiv z'-Z(t')$, and expansion of $Z(t')$ around $Z(t)$, we obtain

\begin{eqnarray}\label{expansion}
&&\mathcal{G}(\zeta-\zeta'+Z(t)-Z(t'),\,t-t')=\\&&\,\nonumber\\
&&\int_{-\infty}^{+\infty}\frac{dk}{2\pi}\exp{[-k^2(t-t')+ik(\zeta-\zeta')+ik(t-t')v(t)]}\nonumber\\
&&\cdot\,\exp{\left[-ik\sum_{n\ge 2}\frac{(-1)^n}{n!}(t-t')^n\partial_t^{\,n-1}v(t)\right]}\nonumber
\end{eqnarray}
where $v(t)\equiv\dot Z(t)=v_P+\dot h(t)$.

If one temporarily applies the scaling transformations $h\rightarrow m^{-2}h,\partial_t\rightarrow m^2\partial_t$, one observes that, whereas $v(t)$ remains unchanged, a factor $m^{2n-2}$ is attached to the contributions $\propto\partial_t^{\,n-1}v(t)$. Therefore, with increasing $n$ these terms are progressively negligible in Eq. (\ref{expansion}) due to the smallness of $m^2$. Neglecting all terms of order $n\ge 2$, we encounter the quasi-stationary approximation, which is often used in phenomenological approaches. As we shall see, however, a proper understanding of oscillatory motions of the interface requires to incorporate the term of order $n=2$. 

Evaluation of Eqs. (\ref{concentration}) - (\ref{expansion}) then leads to the expression

\begin{equation}\label{trapping}
C(z,t)=C_P(\zeta;v)+\,\dot v\,\frac{1}{2}\frac{\partial^2}{\partial v^2}\frac{1}{v}[C_P(\zeta;v)+C_P(\zeta;0)] 
\end{equation}
with $C_P(\zeta;v_P)$ determined by the last line in Eqs. (\ref{balance}). Insertion of this result into the first equation in Eqs. (\ref{integro}) yields the relation  

\begin{eqnarray}\label{differential}
&&\frac{1}{p}\,v=F_P-m^2(Z-Z_P)\\&&+\,G_P(v)-G_P(0)+\dot v\frac{1}{2}\frac{\partial^2}{\partial v^2}\frac{1}{v}[G_P(v)+G_P(0)]\nonumber
\end{eqnarray}
with $G_P(v)$ following from the second line in Eqs. (\ref{balance}). For $v=v_P$ the result (\ref{differential}) consistently reduces to the first equation in Eqs. (\ref{balance}). Subtracting the latter from Eq. (\ref{differential}), we, finally, find for the displacement $h(t)\equiv Z(t)-Z_P(t)$ the differential equation

\begin{equation}\label{oscillator}
M(\dot h(t))\,\ddot h(t)+R(\dot h(t))+m^2\,h(t)=0
\end{equation}
where we have introduced the mass and friction functions

\begin{eqnarray}\label{coefficients}
&&M(\dot h)\equiv -\,\frac{1}{2}\,\frac{\partial^2}{\partial v_P^2}\left[\frac{G_P(v_P+\dot h)+G_P(0)}{v_P+\dot h}\right]\,\,,\nonumber\\
&&R(\dot h)\equiv\frac{1}{p}\,\dot h-G_P(v_P+\dot h)+G_P(v_P)\,\,.
\end{eqnarray}
Eq. (\ref{oscillator}) has the typical appearance of a nonlinear damped oscillator, and represents one of the central results of the present Letter. We mention that, due to the singular dependence of $M(\dot h)$ on $v_P+\dot h$, the differential equation (\ref{oscillator}) is only valid below the crossover line $m^2\propto (v_P+\dot h)^3$.

\begin{center}
\begin{figure}[t]
 \includegraphics[width=8cm]{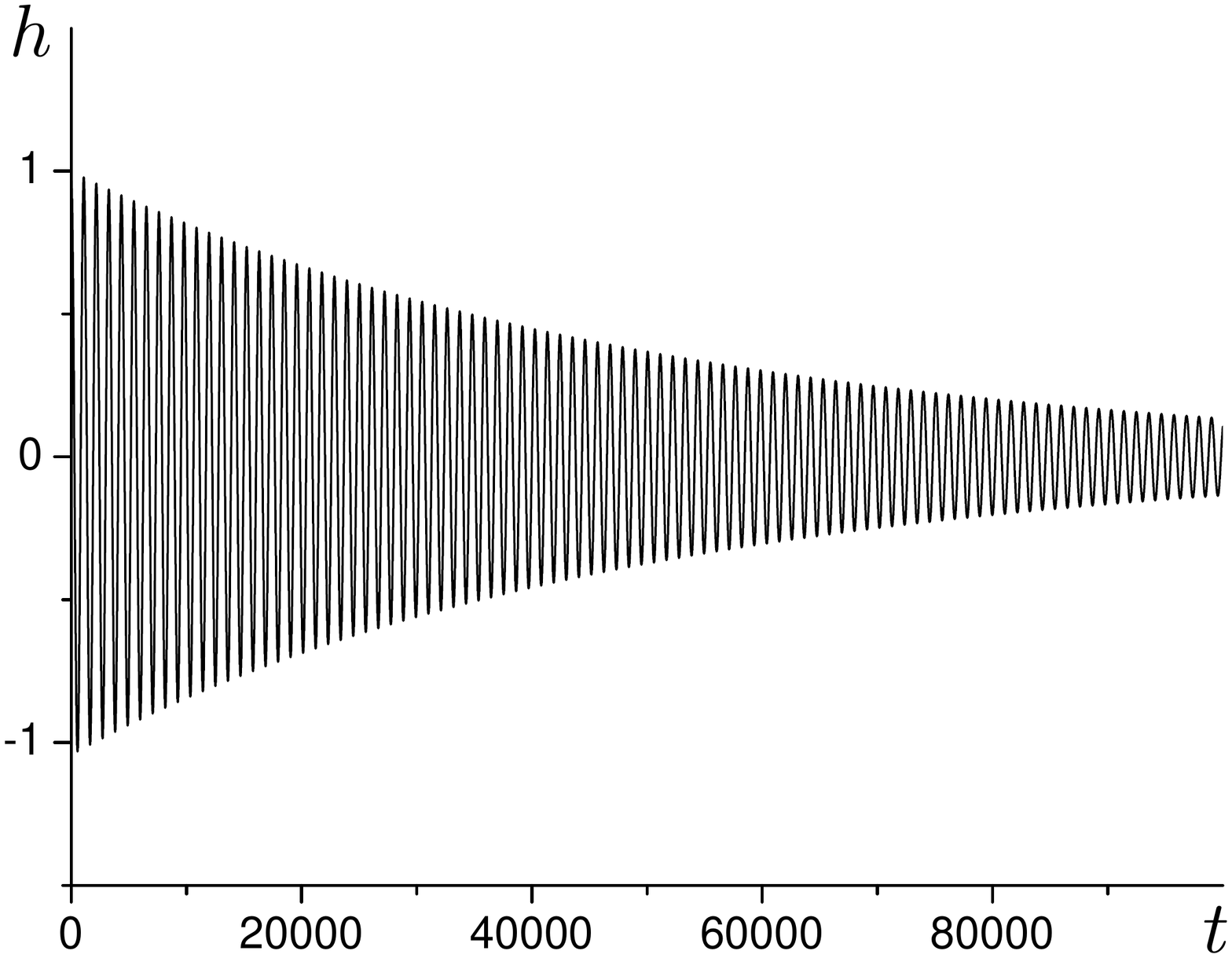}
 \includegraphics[width=8cm]{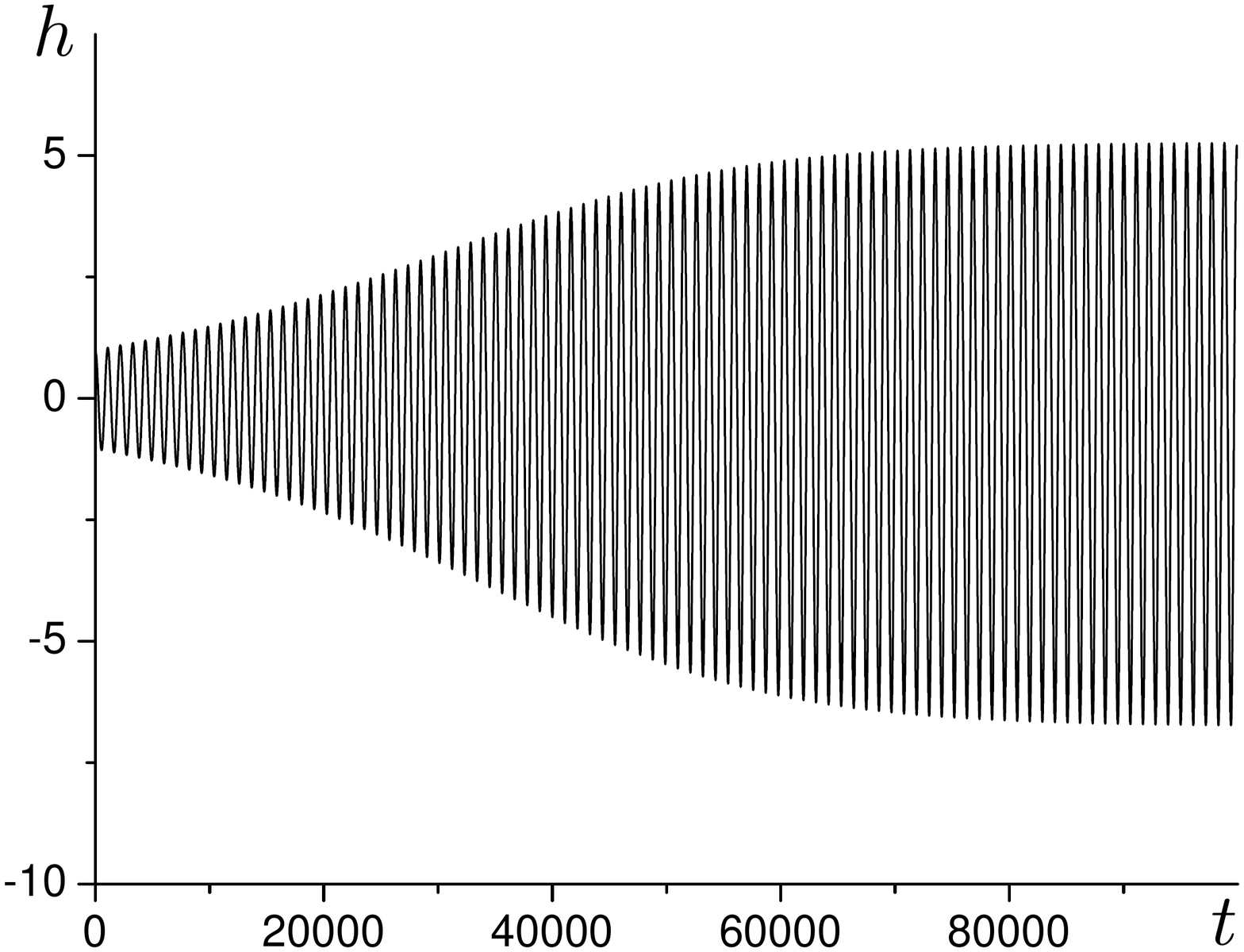}
 \caption{Numerical solutions $h(t)$ for $\gamma=0.01$, $p=100$, $m=0.003$, approaching the value $h=0$ for $v_P=0.522$, and a limit cycle for $v_P=0.520$.}
\end{figure}
\end{center}

In order to check the stability of the obvious solution $h(t)=0$, we linearize Eq. (\ref{oscillator}) in h(t), which, due to the definitions (\ref{coefficients}) and the first line in Eqs. (\ref{balance}), yields   
\begin{equation}\label{linearized}
M(0)\,\ddot h+F_P'(v_P)\dot h+m^2\,h=0\,\,.
\end{equation}
In this ordinary oscillator equation the friction coefficient can change sign at some critical velocity $v_C$, defined by $F_P'(v_C)=1/p-G_P'(v_C)=0$. A similar stability limit has been found by Cahn in grain-boundary motion \cite{Cahn}.

For quantitative discussions of the behavior of $h(t)$ we now adopt the specific model

\begin{equation}\label{kink}
U(\zeta)=\Theta(-\zeta)\exp{\zeta}\\+\Theta(\zeta)[2-\exp{(-\zeta)}]\,\,, 
\end{equation}
derived in Ref. \cite{PRE} from a double-parabola phase-field model. Then, solving the integrals in Eqs. (\ref{balance}), one finds

\begin{equation}\label{drag}
G_P(v)=-\,\gamma\frac{v+2}{(v+1)^2}\,\,,
\end{equation}
which determines all terms in the oscillator equation (\ref{oscillator}).

The resulting numerical solutions for $h(t)$ above and below the Cahn threshold $v_C$ are shown in Fig. 1 where the approach to a limit cycle in the unstable regime is clearly visible. For small distances $\vert v_P-v_C\vert/v_C\ll 1$ the envelopes of these curves can be calculated analytically by the Bogoliubov-Mitropolsky method \cite{Bogoliubov}. To leading order one finds

\begin{equation}\label{B-M}
h(t)=a(t)\cos{\psi(t)}
\end{equation}
where $\psi(t)$ is a rapidly oscillating phase, and $a(t)$ is an amplitude, obeying the differential equation
\begin{equation}\label{amplitude}
\frac{da}{dt}=-\rho_1\,a-\rho_3\,a^3\,\,,
\end{equation}
with $\rho_1\equiv r_1(v_P-v_C)$, and the parameters $r_1,\rho_3$ fixed by the values of $\gamma,p$. The solution of Eq. (\ref{amplitude}) reads

\begin{equation}\label{a-solution}
a(t)=a_0\left\{\left[1+\frac{\rho_3}{\rho_1}a_0^2\right]\exp{[2\rho_1\,t]}
-\frac{\rho_3}{\rho_1}a_0^2\right\}^{-1/2}\,\,,
\end{equation}
which for $\rho_1>0$ and $\rho_1<0$ describes the envelopes in Fig. 1. The asymptotic value of the limit-cycle amplitude shows the critical behavior $a(\infty)=\sqrt{-\rho_1/\rho_3}$.

The numerically obtained limit-cycle trajectories of the quantities $h(t)$, $\dot h(t),C(Z(t),t)$ deep inside the unstable regime are displayed in Fig. 2. They also inform on the local temperature at the oscillating interface, since this is measured by the quantity $m^2h(t)$. The pronounced oscillations of the solute concentration at the interface $C(Z(t),t)$ reflects the appearance of solute bands. From the curves in Fig. 2, which in part are remarkably close to the findings by Conti in Ref. \cite{Conti}, one concludes that the high- and low-concentration layers are connected by large-acceleration segments, explaining the sharpness of the interfaces between these layers.

In order to explore the possible formation of dendritic ripples, one has to consider perturbations of the form $h({\bf x},t)=\hat h({\bf q},\omega)\exp{(i{\bf q}\cdot{\bf x}+\omega t)}$ in a three-dimensional version of the model (\ref{model}). In view of the almost stationary behavior of $\dot h(t)$ in Fig. 2 at low velocities, we choose, as an approximation, $\dot Z(t)=v_P$ as a reference velocity. Following Ref. \cite{PRE}, we then find the dispersion relation

\begin{eqnarray}\label{dispersion}
\frac{\omega}{p}+q^2+m^2-v_P[G_P(v_P+\lambda)-G_P(v_P)]&=&\nonumber\\\frac{\lambda^2-q^2 }{v_P+2\lambda}\, [G_P(v_P+\lambda)+ G_P(\lambda)]\,\,,
\end{eqnarray}

\begin{center}
\begin{figure}[h]
 \includegraphics[width=8cm]{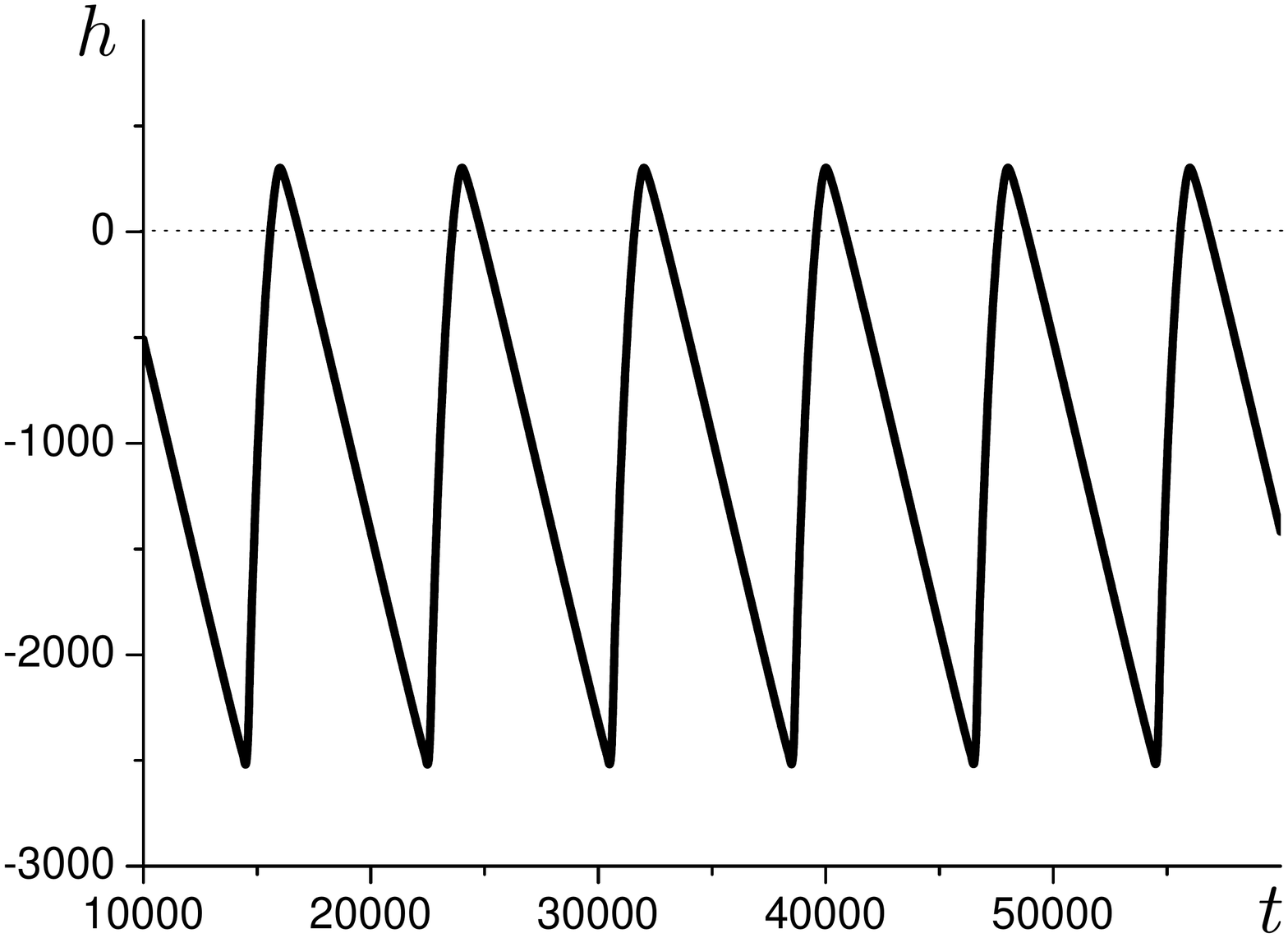}
 \includegraphics[width=8cm]{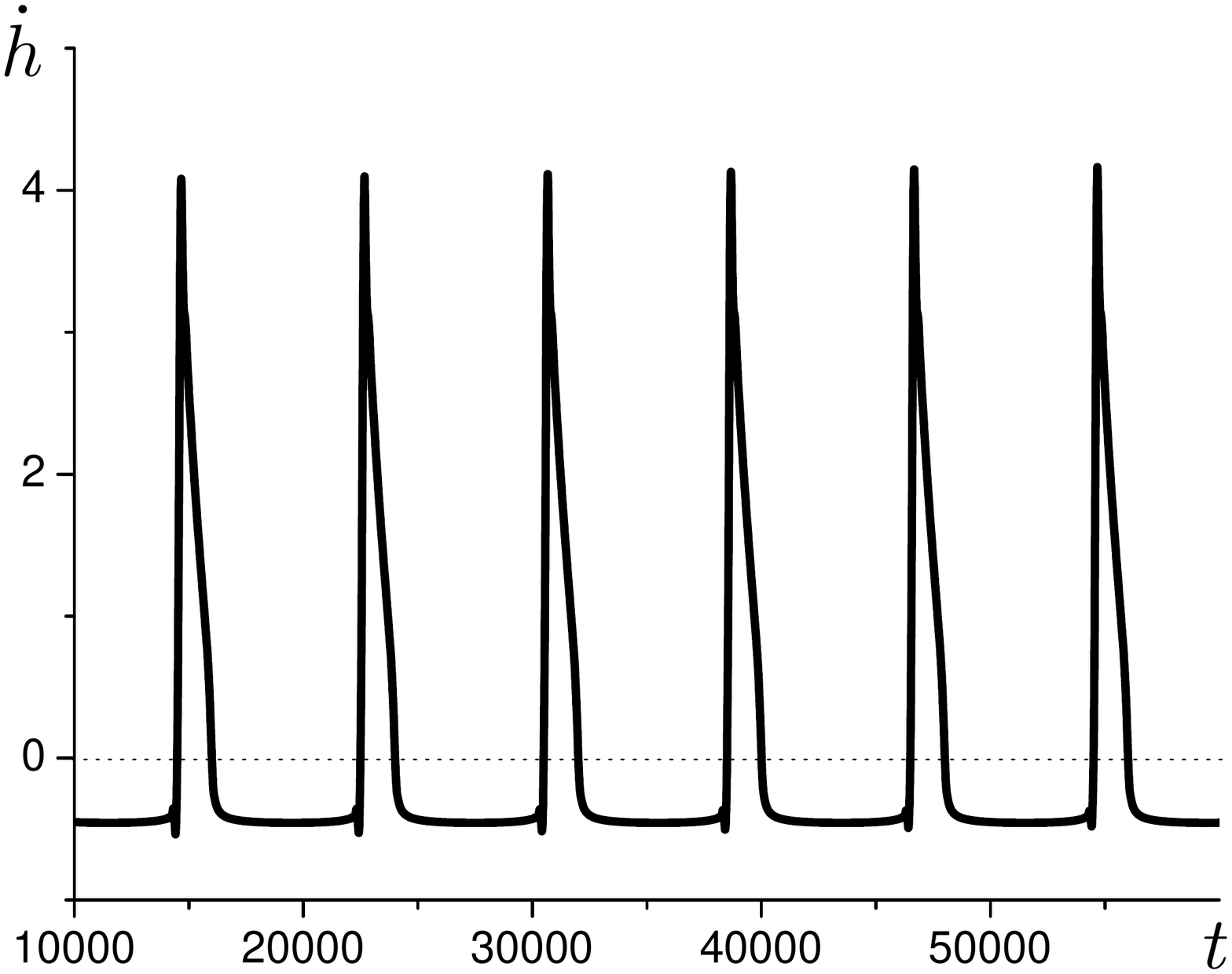}
 \includegraphics[width=8cm]{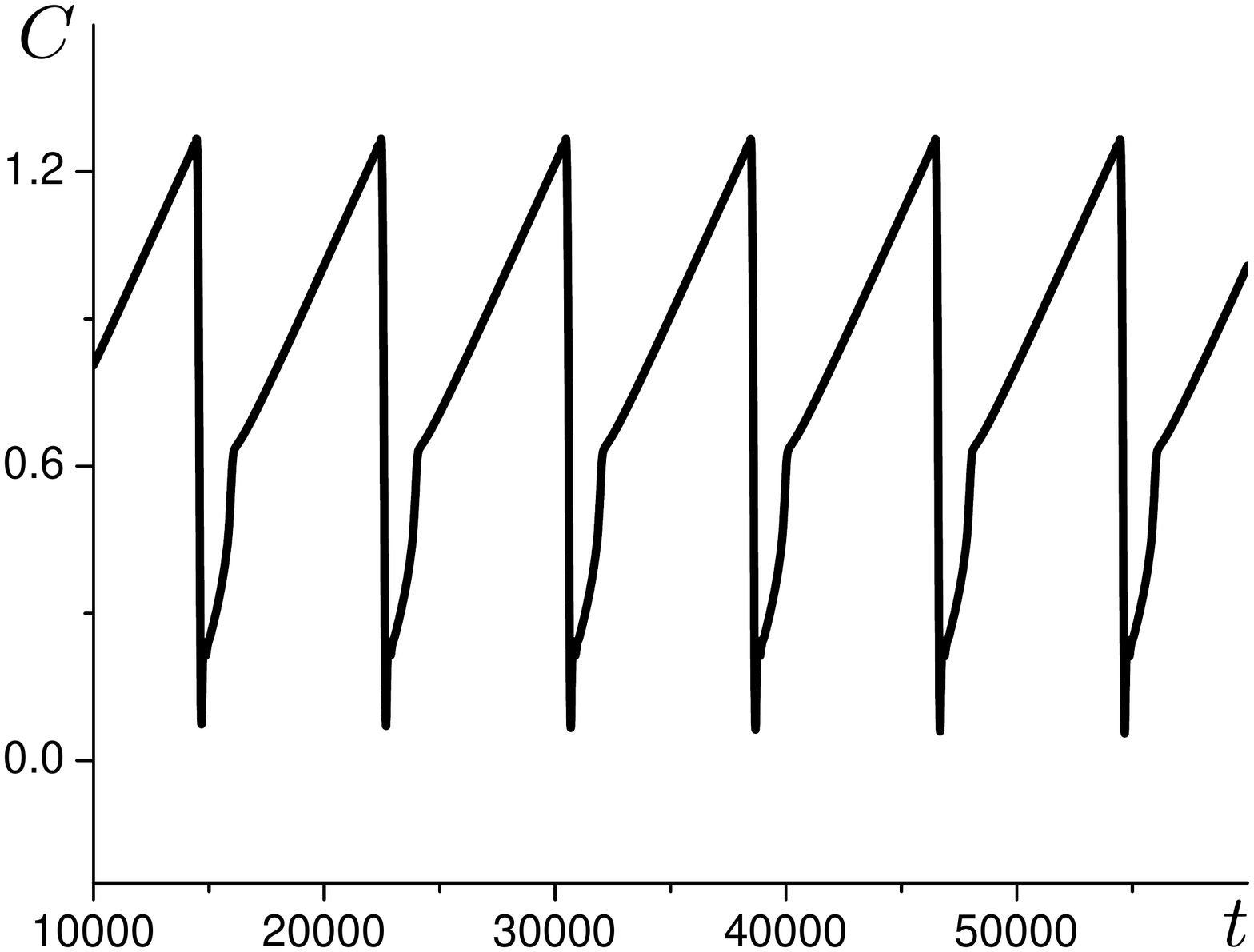}
 \caption{Trajectories of $h(t),\dot h(t),C(Z(t),t)$ in the unstable regime for $\gamma=0.02$, $p=100$, $m=0.003$, and $v_P=0.5$.}
\end{figure}
\end{center}
with $\lambda\equiv -(v_P/2)+\sqrt{(v_P/2)^2+\omega+q^2}$ where the term $m^2$ is the only new element. The wave-number threshold $q_c$ for the Mullins-Sekerka instability \cite{MS} is determined by the relations $\omega_1(q_c)=\omega_1'(q_c)=0$. By elimination of $q_c$ from these equations one generates the neutral-stability boundary of the instability in form of a function $v_P(\gamma)$, with a parametric dependence on $m$. 

In Fig. 3 the projection of the limit cycle, belonging to Fig. 2, enters the Mullins-Sekerka unstable regime at low velocities where the interface develops a dendritic microstructure, a typical feature of banded structures in metallic alloys. The other small cycle in Fig. 3 generates layers of precipitation-free    periodic solute concentrations. 

The most obvious generalization of our procedure is to explore the formation of non-planar layering effects, which also is a field for experimental investigations. A typical example of such an effect is the oscillatory growth of a spherical nucleus, which has been discussed on the basis of a phase-field model in Ref. \cite{KCGOSI}, and which we are going to reconsider within our approach. 

\begin{center}
\begin{figure}[t]
 \includegraphics[width=8cm]{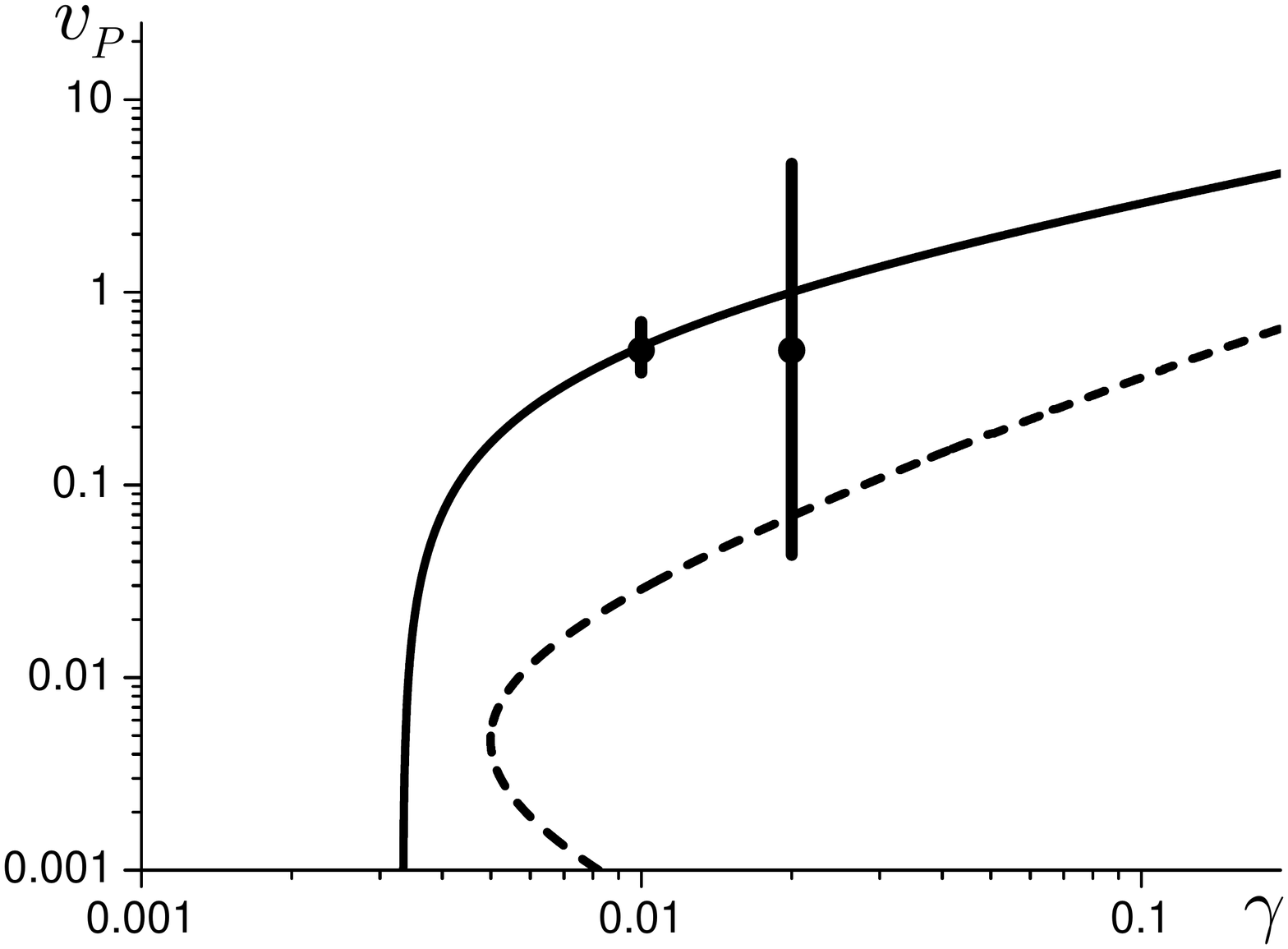}
 \caption{Neutral stability curves, enclosing the regions of the Cahn (solid line), and of the Mullins-Sekerka instability (dashed line). The vertical lines are projections of limit cycles at $\gamma=0.01$ and $\gamma=0.02$, both at $p=100,m=0.003$ and $v_P=0.5$.}
\end{figure}
\end{center}

A. L. Korzhenevskii wants to express his gratitude to the University of D\"usseldorf for its warm hospitality. This work has been supported by the DFG under BA 944/3-3, and by the RFBR under N10-02-91332.

\end{document}